\documentclass[letterpaper,twocolumn,10pt]{article}
\usepackage{usenix,epsfig,endnotes}
\usepackage{graphicx}
\usepackage{tikz}
\newcommand*\circled[1]{\tikz[baseline=(char.base)]{
    \node[shape=circle,draw,inner sep=1 pt] (char) {#1};}}
\graphicspath{ {./image/} }

\begin{document}

%don't want date printed
\date{}

%make title bold and 14 pt font (Latex default is non-bold, 16 pt)
\title{\Large \bf Hibernate Container: A Deflated Container Mode for Fast Startup and High-density Deployment in Serverless Computing}

\author{
{\rm Yulin Sun} \\
Quarksoft
\and
{\rm Deepak Vij} \\
Futurewei
\and
{\rm Fenge Li} \\
Futurewei
\and
{\rm Wenjian Guo} \\
Futurewei
\and
{\rm Ying Xiong} \\
Futurewei
}

\maketitle

% Use the following at camera-ready time to suppress page numbers.
% Comment it out when you first submit the paper for review.

\subsection*{Abstract}
Serverless computing is a popular cloud computing paradigm, which requires low response latency to handle on-demand user requests. There are two prominent techniques employed for reducing the response latency: keep fully initialized containers alive (Warm Container\footnote{Warm Container refers to fully initialized container created as part of warm start process}) or reduce the new container startup (cold start) latency.

This paper presents the 3rd container startup mode: Hibernate Container\footnote{Hibernate Container refers to Hibernated Container Mode}, which starts faster than the cold start container mode and consumes less memory than the Warm Container mode. Hibernate Container is essentially a ”deflated” Warm Container. Its application memory is swapped out to disk, the freed memory is reclaimed and file based mmap memory is cleaned-up. The Hibernate Container’s deflated memory is inflated in response to user requests. As Hibernate Container's application is fully initialized, its response latency is less than the cold start mode; and as the application memory is deflated, its memory consumption is less than the Warm Container mode. Additionally, when a Hibernate Container is ”woken up” to process a request, the Woken-up Container has similar response latency to Warm Container but less memory consumption because not all the deflated memory needs to be inflated. We implemented the Hibernate technique as part of the open source Quark secure container runtime project and our test demonstrated that Hibernate Container consumes about 7\% to 25\% of the Warm Container memory.  All of this results in a higher deployment density, lower latency and appreciable improvements in the overall system performance.

\section{Introduction}

Serverless computing, including Function as a Service (FaaS) and Serverless container, is more and more becoming a popular cloud computing paradigm. Serverless computing environment typically runs multi-tenant workloads in a shared environment to handle on-demand user requests. It has been supported by major cloud providers such as AWS Lambda \cite{lambda}/Fargate \cite{Fargate}, Google Function \cite{googlefunction}/Cloud Run \cite{cloudrun}, Azure Function \cite{azurefunction}/Container Instance \cite{azurecontainer}.

In order to host multi-tenant user application workloads in a shared environment, major cloud providers generally use VM based secure container runtime instead of common process based container runtime like runC/LXC. The VM based secure container runtime provides security isolation just like a conventional virtual machine. For example, AWS uses Firecracker\cite{FirecrackerRecommand}, GCP uses gVisor\cite{gVisor} and AliCloud and HuaweiCloud use Kata containers\cite{kata} for serverless computing. But these VM based secure container runtimes consume more memory and have higher response latency than the process based container runtime.

The response latency for a user request is utmost important for serverless computing environment. The response latency for a user request includes 3 parts: container runtime startup, user application initialization and user request processing. The container runtime startup time typically takes 100 or so milliseconds and application initialization time ranges from 10 milliseconds to 10 seconds, while the user requests process time is typically short with a few milliseconds to 10 seconds. Compared with the request process time, the container runtime startup and the application initialization latency contribution is quite substantial.

Reducing the container runtime startup and application initialization latency is one of the key challenges of serverless computing design. There are generally 2 optimizations employed to reduce the latency:
\begin{itemize}
\item \textbf{Warm startup optimization}: A common technique is to keep the execution runtime alive, i.e. Warm Container, for a short period, so that the future invocation of the same request can reuse the Warm Container. The warm startup does reduce the cold start overheads but keeping a container alive consumes substantial computer resources, especially memory resources. This, in turn, results in higher system resource requirements. There are various ongoing efforts for improving the warm startup efficiencies such as reducing container runtime overhead\cite{RunD}\cite{gVisor} and optimizing warm container keep-alive schedule policy\cite{faascache}.
\item \textbf{Cold startup latency optimization}: Of course, we can’t afford to keep all the serverless containers alive because of system resource limitations. Another research area is to reduce the cold startup latency. This effort includes reducing the container runtime startup time\cite{RunD}\cite{gVisor} and application initialization latency\cite{sock}\cite{catalyzer}\cite{RCEP}.
\end{itemize}

This paper presents the 3rd startup mechanism based on the notion that the conventional memory swapping technique is quite suitable for mitigating the Serverless computing startup time. Following are the key considerations to this regards: 
\begin{itemize}
\item \textbf{Fast swapping storage}:   With the advent of high performance secondary storage such as SSD, NVM commercially availability in public cloud \cite{optane}, memory swapping performance gains huge improvements.

\item \textbf{Lightweight serverless workload}: The fast startup requirements of serverless computing environment requires a lightweight and low memory footprint workloads. For example, in AWS \cite{serverlesslambda}, \textit{47\% of functions are configured to run with the default minimum memory setting of 128 MB. Overall, only 14\% of AWS lambda functions have a memory allocation greater than 512 MB}. And in Azure\cite{serverless}, \textit{90\% of the applications never consume more than 400MB, and 50\% of the Serverless application workloads allocate at most 170MB}. As the Serverless computing workload memory footprint is small, the resulting swapping memory cost incurred is quite low.
\end{itemize}

In summary, there is a great opportunity to adopt the memory swapping technique in the Serverless computing environment for achieving low latency startup in conjunction with low memory consumption of a keep-alive container. With memory swapping technique as the key enabler, our paper proposes and implements Hibernate Container, a deflated keep-alive Warm Container. Hibernate Container leverages the following key optimizations for achieving the low latency startup and low memory consumption:
\begin{itemize}
\item \textbf{Memory}: Memory consumption of Hibernate Container is much lower than the Warm Container as it swaps out the application memory to disk, reclaims and returns the container application free memory to the host OS kernel, and finally cleans up the file backed mmap memory and returns to host OS.
\item \textbf{CPU}: It doesn't consume system CPU cycles at all as it puts the user applications in a complete pause state.
\end{itemize}

Hibernate Container user request response latency is much lower than the cold startup. It is mainly because the user application of Hibernate Container are fully initialized and the container runtime is running with the following keep-alive resources:
\begin{itemize}
\item \textbf{Host OS objects}: Hibernate container keeps its host OS objects alive, such as container runtime OS process, Cgroup, container network, container file system, processes. The OS objects consume little system memory but keeping them alive saves much re-initialization cost.
\item \textbf{Blocked container runtime threads}: Container runtime’s host thread is blocked to wait for user request. It doesn’t consume CPU cycles but the system immediately responds similar to a Warm Container. 
\end{itemize}

It is important to highlight that a Woken-up Container response latency for subsequent requests is almost similar to a Warm Container but it consumes less memory than Warm Container. This is mainly because not all the deflated memory is needed and inflated by a Woken-up container for user request processing. 

Overall, all of this results in a higher deployment density and better system performance as a whole.

The following is a summary of our key contributions:
\begin{itemize}
\item We propose and implement Hibernate Container mode as part of the open source Quark container runtime \cite{quark}. Hibernate Container mode consumes less memory than a Warm Container and starts faster than the cold startup. Additionally, Woken-up Container derived from the Hibernate Container consumes less memory than a Warm Container and has almost similar request response latency.
\item We identify that the major memory swap-in latency is due to random read of the SSD disk. Inspired by the REAP[16], a record-and-prefetch mechanism, we implement the batch memory pre-fetch swap-in mechanism as part of the Hibernate Container inflate process. We compare the page fault based swap-in and REAP swap-in with different benchmarks.
\item We implement a new reclaim oriented memory management algorithm for returning free memory pages efficiently to the host OS kernel. This avoids need for a complex Ballooning technique.
\end{itemize}

\section{Background and Motivation}

Hibernate Container is a deflated Warm Container implemented in Quark Container runtime. Its deflation includes reclaiming user application free memory and swapping out user memory to secondary storage.

This section discusses the background of 
% overall Serverless platform architecture, 
the Quark secure container runtime design,  the existing guest OS freed memory reclamation and swapping mechanisms. Finally, this section discusses motivation for our paper and enumerates various opportunities for optimizing the contemporary state-of-the-art Serverless computing environment.

% \subsection{Typical Serverless Computing Platform}
% \includegraphics[scale=0.4]{serverless_arch}
% \begin{center}
%  Figure 1: Overall architecture of a Serverless platform
% \end{center}

% % Serverless computing includes FaaS, e.g. AWS Lambda\cite{lambda} and serverless container, e.g. AWS Fargate\cite{Fargate}. 

% Figure 1 illustrates a typical architecture of Serverless computing platform. The frontend accepts user requests such as HTTP request, event trigger. The user request is processed by the containerized workloads running on the Worker Nodes. When the frontend gets a user request, it asks for a user request handler container from the Controller. The Controller returns a keep-alive container or cold starts a new container. The Controller manages the container startup, keep-alive or eviction. The Controller manages the container startup and eviction through the Node Agent running on the Worker Node. After the frontend gets the user request handler container, it forwards the request to the container using a Messaging Bus.
\subsection{Secure Container and Quark Runtime}
As mentioned earlier that we implemented Hibernate Container mode as part of the open source Quark container runtime \cite{quark}. In this sub-section, we briefly describe various state-of-the-art secure container runtime technology landscape. We specifically delve deep into details of Quark runtime architecture.

Serverless computing hosts multi-tenant workloads in a shared environment. Conventional container runtimes such as RunC/LXC are not suitable for Serverless computing environment. This is primarily because these container runtimes can't provide multi-tenant level secured isolation. Instead, major cloud providers are using Virtual Machine (VM) level secure container runtimes as part of their respective Serverless computing environment. Examples of such VM level container runtimes are Kata\cite{kata}/Firecracker\cite{firecracker} and gVsior\cite{gVisor}. 

\includegraphics[scale=0.35]{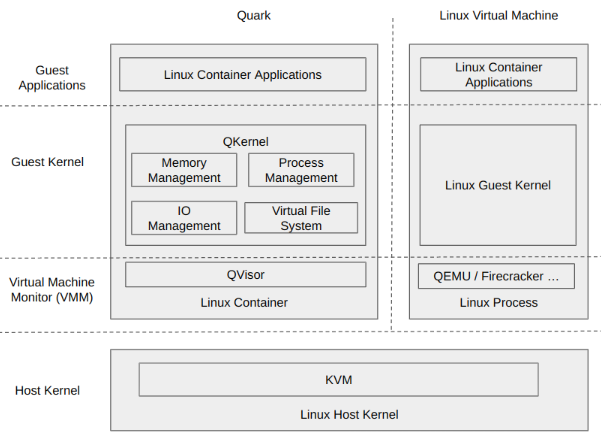}
\begin{center}
 Figure 2: Quark Runtime Architecture
\end{center}
Kata and Firecracker use Linux kernel based Virtual Machine for secure isolation. As both of them are using the existing general purpose Linux kernel, their startup latency and resource overheads are quite high for Serverless computing environment.

Quark\cite{quark} and gVisor\cite{gVisor} are two other notable secure container runtimes designed specifically for Serverless computing. They consist of a userspace OS Kernel and a lightweight Virtual Machine Monitor (VMM). They are designed to provide a Linux compatible system call interface as well as provide support for CRI/OCI interface so that existing Linux container images can run without any changes. As mentioned earlier that these two secure container runtimes are highly optimized for serverless workloads. As a result, their startup latency and resource overhead is lower than Kata/Firecracker runtimes. 
% Furthermore, Quark provides seamless support for Serverless computing specific features such as support for Hibernate Container mode, which is hard to implement in the Linux Kernel. 
However, unlike Kata/Firecracker secure container runtimes which are explicitly based on Linux Kernel, Linux compatibility for Quark and gVisor secure container runtimes is not as good.

Figure 2 illustrates Quark’s architecture. Quark has similar architecture to conventional Linux Virtual Machine does. They both run on top of Linux Host kernel with KVM Hypervisor. Quark runtime process runs within a standard Linux container with Cgroup and network/file system namespace for isolation. Quark includes a new userspace OS Kernel (QKernel) and VMM (QVisor) heavily optimized for Severless computing. Quark provides support for a virtualized system call interface which simulates Linux system calls. Quark implements majority of Linux Kernel functions such as memory management, process management and IO management etc. 

Quark is specifically designed for Serverless computing environment, and it is to integrate Serverless computing specific features such as support for Hibernate Container mode etc.

\subsection{Guest OS Freed-up Memory Reclamation}
As mentioned earlier that one of the key value proposition of Hibernate Container mode is to return user application freed-up memory to the host OS.

Freed-up memory reclamation process is not a trivial task for a general purpose guest OS such as Linux. Typically, when a Linux guest OS’s application frees-up memory to the Linux Guest OS kernel, ideally the Linux Guest OS kernel should in turn return the application's freed-up memory back to the host Linux kernel so that freed-up memory may be re-allocated to other host OS processes. Unfortunately, Linux Guest OS keeps the freed memory in its memory pool and doesn't return it to the host Linux OS. It is because Linux is optimized for bare metal machine environment in which there is no memory reclamation requirements. In a nut-shell, what this means is that memory freed-up by guest OS is not released and reclaimed back by the host OS in a conventional virtualization setup.  

Currently, there are two distinct approaches to address Linux guest OS freed memory reclamation issue:
\begin{itemize}
\item \textbf{Ballooning}\cite{ballooning}: Ballooning approach relies on a special ballooning driver that resides in the guest OS and cooperates with the hypervisor to adjust a VM’s memory size dynamically. Essentially in a nut-shell, Ballooning is a computer memory reclamation technique used by a hypervisor to allow the physical host system to retrieve unused memory from certain guest VMs and share it with others.
\item \textbf{Memory plugin}\cite{hotplug}: Memory plugin mechanism relies on OS Kernel support of hot-add/hot-remove physical memory. Memory hot-remove makes memory sections unavailable for users. It needs to do Page migration to move the used pages to another section. However, Page migration causes a performance penalty. Memory plugin has been used in VM memory harvesting \cite{harvesting}. 
\end{itemize}

VM based container runtimes such as Kata/Firecracker are also based Linux guest OS so they also suffer the freed-up memory reclamation issue. Based on our knowledge, neither Ballooning nor Memory plugin approaches have been adopted by them. This is mainly due to the fact that both Ballooning and Memory plugin approaches are too complex to be adopted in the Serverless computing environment. 

As part of our Hibernate Container work, we have implemented a dedicated memory management system for improving the reclamation efficiencies in Quark secure container runtime for the Serverless computing environment.

\subsection{Guest Application Memory Swapping}
As mentioned earlier that one of the key value proposition of Hibernate Container mode is to swap out user application memory.

Swapping is a memory management scheme that temporarily swaps out the inactive memory pages to a secondary storage and marks the page’s entry in the page table as Not Present. When the system needs to access the swapped-out page, the OS virtual memory management system generates a page fault in order to swap-in the page.

For a common virtualization environment, the current host OS swapping mechanism is not efficient because currently it is done in an uncooperative manner. VSWAPPER\cite{vswapper} work has investigated the uncooperative swapping inefficiencies in a virtualization environment. The issues include Silent swap writes, Stale swap reads and False swap reads etc. VSWAPPER implements a guest agnostic memory swapper to address these all these issues. VSWAPPER work is quite good for addressing the uncooperative swapping issue in a common virtualization environment but not specifically for Serverless computing environment. 

For Serverless computing, we have numerous opportunities to achieve better swapping performance as we would like to swap out the entire memory of an idle container:

\begin{itemize}
\item \textbf{Batch swap-out of all the application memory pages}: For common swapping scenarios, the OS kernel chooses to swap-out inactive pages. In a Serverless computing environment, we swap-out all the user application memory set of an idle container. This, in turn, resulting in substantial memory management process related cost savings.
\item \textbf{Race condition free Swap-out for paused applications}: In the Serverless computing environment, we can pause the idle user application processes while swapping-out memory pages. This avoids complex race condition handling of common swapping process.
\item \textbf{Batch sequential disk read for page swap-in}: A common OS memory page swap-in is triggered by a page fault and the secondary swapping storage is accessed with random read. For the common secondary storage, no matter HDD or SSD, batch sequential reads always have much better performance than a random read. REAP\cite{RCEP} reveals that functions access the same stable working set of pages across different invocations of the same function. After identifying the set of pages, these pages can be prefetched using batch sequential reads. Compared with page fault based swap-in approach, batch swap-in approach can not only result in savings related to disk random page load cost but also savings related to the cost of page fault handling and switching between guest and host.
\end{itemize}

Our motivation is to address all of the above mentioned optimization opportunities and develop a much more efficient swapping mechanism for the Serverless computing environment.

\section{Design and Implementation}
The Following subsections delve deep into architectural and design details for enabling Hibernate container functionality as part of Quark secure container runtime.

\subsection{Container State Machine with Hibernate Container}
This section introduces how Hibernate Container works in response to an incoming user request.

Figure 3 shows various container state transitions for serving the incoming user requests.

\includegraphics[scale=0.35]{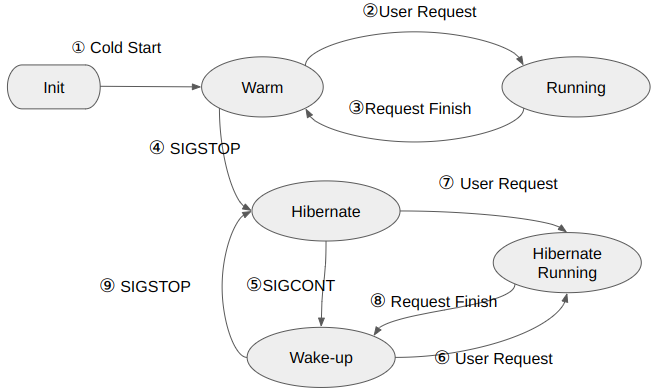}
\begin{center}
 Figure 3: Container State Machine
\end{center}

At the time of an incoming user request, Serverless Platform performs \circled{1} a Cold Start. This, in turn, spawns a new Warm Container, the user request is forwarded to the newly spawned Warm Container. After the Warm Container receives the user request, it \circled{2} transitions into the Running state to process the request and when the request process is finished, it \circled{3} returns back to the Warm state. 

In order to reduce the user request response latency, the Serverless Platform may keep the Warm Container alive for a short period of time. In case there are other incoming user requests in a quick succession (following one another at short intervals), these requests may be served by the Warm Container in order to achieve low response latency. However, when the Warm Container stays idle, it keeps on consuming the application specific allocated memory footprint. In that case, the Serverless Platform may evict the Warm Container whenever there is memory pressure to free-up memory for other Serverless function containers. After the Warm Container is evicted, the next user request experiences higher cold start latency. In a nut-shell, more the number of Warm Containers in the system, better the user request response latency is. 

In addition to the above mentioned container states in the conventional Serverless computing environment, we propose 3 new container states as below: 

\textbf{Hibernate}: Hibernate Container is a deflated Warm Container which consumes less memory than Warm Container. Serverless Platform may choose to "deflate" a Warm Container to Hibernate Container instead of evicting the Warm Container altogether in order to free-up memory. Serverless Platform may initiate deflation of a Warm Container \circled{4} to Hibernate Container by sending a SIGSTOP signal to the Warm container.

\textbf{Hibernate Running}:  Hibernate Container may \circled{7} transition to the Hibernate Running state when it gets a user request to process the user request just like Running Container does. 

\textbf{Woken-up}: Hibernate Running container \circled{8} returns to Woken-up state when the user request process finishes. The Woken-up Container \circled{6} enters Hibernate Running state again at the time of a next user request. A Woken-up Container may also \circled{9} return to the Hibernate state when it receives SIGSTOP signal.  A Woken-up Container user request response latency is almost similar to the Warm Container but consumes less memory. When Serverless Platform predicts there is an incoming user request, it may also \circled{5} "wake up" a Hibernate Container to a Woken-up Container in anticipation by sending a SIGCONT signal to the Hibernate Container in order to reduce the response latency.

\subsection{Deflation Process Overview}

Hibernate container is essentially a deflated Warm Container. Hibernate Container is derived from Warm Container through the following 4 steps:
\begin{enumerate}
\item Pause Warm Container user application processes and block the container runtime host OS threads to wait for "wake-up" trigger.
\item Reclaim and return freed application memory pages to host Linux kernel. 
\item Swap-out application committed memory pages to local disk.
\item Clean up application’s file backed mmap memory through madvise() with MADV\_DONTNEED as “advice” argument \cite{madvise} and return them to Linux kernel.
\end{enumerate}
With step\#1, Hibernate Container does not consume any system CPU cycles. With step\#2 \#3 \#4, Hibernate Container’s user application allocated memory is returned to the host Linux kernel so that it consumes much less memory than a Warm Container. We describe details for steps \#2 \#3 \#4 in subsections 3.3, 3.4, 3.5 respectively.

Hibernate Container may transition into a Warm Container and the transition process includes inflate memory and un-pause the user application processes. Hibernate Container may be woken up by 2 types of triggers:
\begin{itemize}
\item  \textbf{Upon a user request}: When the Serverless Platform gets a user request, Serverless Platform may forward the request to a Hibernate Container directly without waking it up at first in order to reduce the overall system latency. Hibernate Container achieves this by putting the container runtime threads in a blocking state to wait for user requests such as Posix socket sys\_accept or sys\_read. When there is a Posix socket client connect request or the socket data is available, the container runtime thread is unlocked by Linux host kernel and it starts the remaining wake-up processing, i.e., memory swap-in step and the user application process resumes.
\item  \textbf{Serverless system control plane}: Serverless Platform may explicitly wake up a container in anticipation if the Serverless Platform predicts that there will be a user request coming in. Because the memory inflation is partially done before the user request comes in, the user request response latency is lower versus the user request trigger step.
\end{itemize}

\subsection{Memory Reclamation Orientation Memory Management}

Hibernate Container reclaims the guest user application’s freed memory pages and returns it to the host Linux kernel just as VM ballooning technique does. 

QKernel is running in a KVM virtual machine and its guest physical memory is the virtual memory of the host Linux OS. The guest physical memory page (i.e. host virtual memory page) of the QKernel is not committed by the host Linux OS kernel until it is accessed. Quark can return  committed memory pages to the host Linux kernel through syscall sys\_madvise() with MADV\_DONTNEED as “advice” argument \cite{madvise}. After a successful madvise() operation, the subsequent accesses of pages in the range succeeds, but this results in zero-fill-on-demand pages for anonymous private mappings. Ideally, after the free memory regions are identified from the guest OS kernel memory management system, they can be reclaimed by calling madvise().

Unfortunately, it is not easy for original Quark runtime to reclaim freed memory. Quark container runtime currently uses binary buddy allocator \cite{faststorage}. It is not suitable for memory reclamation as it maintains the free memory blocks in a free list, which is a linear linked list and its “next” pointer is kept in the free memory blocks. The free list works fine for a bare metal optimized OS kernel. But for a guest OS kernel, if we use madvise() to reclaim the free memory page blocks, as the subsequent accesses of pages of the memory block is zero-filled, the “next” point is cleared so the free list data structure is broken. In summary, the existing buddy allocator is not suitable for a memory page reclamation. Instead, we implement Bitmap Page Allocator to address this particular issue. We introduce the Bitmap Page Allocator design in the subsequent paragraphs.

Memory allocation for the original Quark container runtime falls into the following 2 areas:

\noindent \textbf{1. Guest user application memory address space allocator}: Guest user applications allocate memory from the guest OS kernel using syscalls such as sys\_brk, sys\_mmap. The syscalls only allocates a memory address space and the memory pages are not be committed until they are accessed via memory page fault handler.

\noindent \textbf{2. Quark runtime global heap allocator}: Quark runtime is developed using Rust programming language. Rust runtime supports a customized heap allocator. Quark uses binary buddy allocator based customized heap allocator. QKernel’s kernel data structures such as kernel stack are allocated from the heap. Original Quark also allocates memory page in page fault handler for user application from the global heap which is not friendly for memory reclamation. As a result of all this, we decided to develop the 3rd memory page allocator for Hibernate Container: Bitmap Page Allocator.

\includegraphics[scale=0.35]{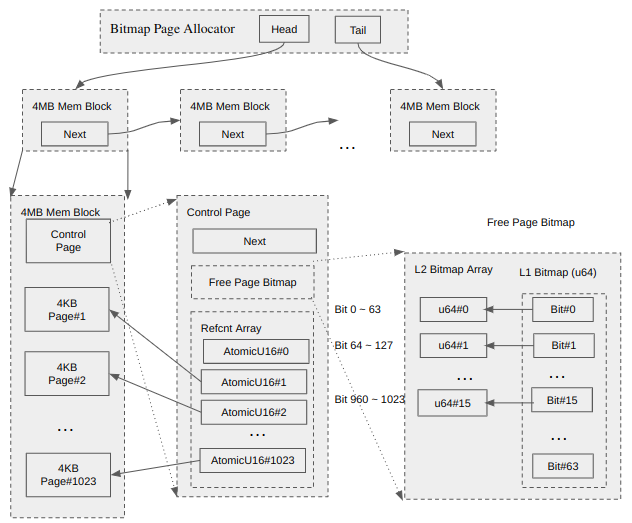}
\begin{center}
Figure 4: Bitmap Page Allocator
\end{center}

The Bitmap Page Allocator is designed for the Guest user application memory page allocation management. It is only used for the fixed size 4KB memory page allocation in page fault handler for user application.

As shown in Figure 4, the Bitmap Page Allocator uses 4MB memory blocks for 4KB memory page allocation. The 4MB memory block’s start address is also 4MB aligned and the first 4KB memory page of the 4MB memory block is reserved as the Control Page. 

The control page contains 3 fields:

\begin{itemize}
\item \textbf{“Next” Pointer}: In the Bitmap Page Allocator, all the 4MB memory blocks with freed pages are linked in a free list which is a linear linked list. The linear linked list “Next” point is kept in the control page.
\item \textbf{Free page bitmap}: One 4MB block contains a total 1024 x 4KB memory pages. Out of this, the first page is used as the Control Page. Remaining 1023 pages can be allocated. The Bitmap Page Allocator keeps a L2 (Level\#2) bitmap with an array of 16 x 64 bits integers (total 1024 bits) to indicate whether the page is allocated or not (i.e. one page with one bit). To accelerate the free page lookup, the Bitmap Page Allocator keeps another 64 bits integer as L1 (Level\#1) bitmap to indicate whether the 64 bits integer in L2 bitmap is zero. So for each page allocation, the Bitmap Page Allocator needs to access two 64 bit integers: one is the L1 bitmap and the other from the L2 bitmap. For example, to find a free page, Bitmap Page Allocator firstly looks for the first non-zero bit of the L1 bitmap 64 bit integer. Provide the first non-zero bit is the 4th bit, it means there is free page indicated by the 4th L2 bitmap 64 bit integer. Bitmap Page Allocator secondly looks for the first non-zero bit of the 4th L2 bitmap 64 bit integer. In this way, the Bitmap Page Allocator free page lookup complexity is O(2).
\item \textbf{Memory page reference count array}: The OS kernel’s memory page may be referenced by multiple page tables when doing process clones. The Bitmap Page Allocator also keeps the page reference count in the control page with an array of 16 bit atomic integers \cite{Atomic} to maximize the memory usage efficeny and improve performance. 
\end{itemize}

Memory page allocation and reference count increases and decreases as below.
\begin{enumerate}
\item \textbf{Memory page allocation}: At the time of a page allocation request, the Bitmap Page Allocator allocates a page from the first 4MB memory block of the free list and update the control page's bitmap. If there is no more free page in the first 4MB memory block, it gets removed from the free list. If there are no more 4MB memory blocks in the free list, the Bitmap Page Allocator allocates another 4MB memory block from the global heap, i.e. the global binary buddy allocator. The memory allocation needs to take a global lock to avoid race conditions.
\item \textbf{Page reference count increase and decrease}: Whenever there is a guest process clone/termination or page copy-on-write (COW), there is guest page reference count increase and decrease. Bitmap Page Allocator also store the page reference count in the control page to improve performance. As the 4MB memory block is 4MB aligned, any guest page may find its Control Page (i.e. the first page of 4MB block) by clearing its address's least 22 bit. So the look up from any user memory page address to its control page doesn’t need a lookup table. After finding the page's Control Page, the reference count increase and decrease is done through Rust’s atomic operation: atomic\_fetch\_add and  atomic\_fetch\_sub \cite{Atomic} which is lockless operation. When the reference count is decreased to zero, the page is freed and returned to Page Allocator by updating the Free Page Bitmap. If the 4MB memory block’s free page count was zero when there is a new free page, the memory block is put back to the free list. When the free page count of the 4MB block reaches max free page count, i.e. 1023, the 4MB memory block can be returned to the global heap.
\end{enumerate}

At the time when the Quark runtime initiates processing to Hibernate a container, it needs to return freed pages to the host Linux kernel. Because the freed data page in the Bitmap Page Allocator is indicated by the bitmap in the Control Page, there is no data stored in the memory page such as "next" point in free list of binary buddy allocator. When doing hibernation, Quark runtime returns the free pages to host OS by calling the host syscall madvise(). In this way, Quark’s hibernate process is much simpler than the VM Ballooning technique. 

Additionally, whenever the Hibernate Container is woken up and the reclaimed memory page is re-allocated to the guest application, the memory page is committed by the host Linux kernel through the host OS page fault. The process is handled by host Linux and is transparent to guest OS Quark. So Quark doesn't need to explicitly re-allocate the page. It, in turn, decreases the wake-up latency and the overall system complexity.

\subsection{Hibernate Container Memory Swapping}
\includegraphics[scale=0.35]{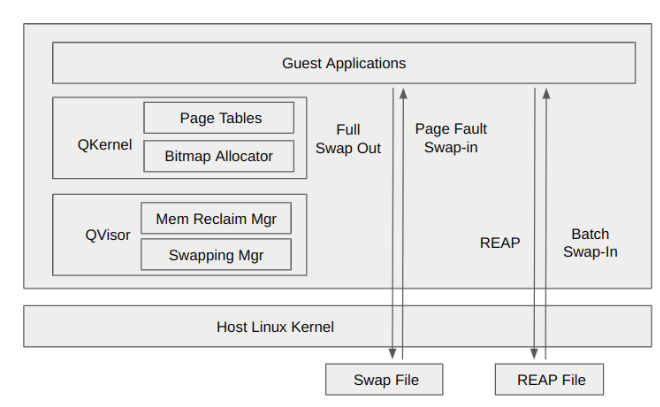}
\begin{center}
 Figure 5: Hibernate Container Swapping
 \end{center}
Hibernate Container swaps out guest application memory to secondary storage and the swapped-out memory pages may be swapped-in at the time when the Hibernate Container is woken up. 

There are 2 types of swap-in mechanisms:
\begin{itemize}
\item \textbf{Page fault based swap-in}: Just like common OS swap-in, when the guest application accesses a swap-out page, a page fault is triggered and the page fault handler loads the memory page.
\item \textbf{Batch REAP swap-in}: A hibernate container may prefetch all the pages recorded by the REAP process.
\end{itemize}

As shown in Figure 5, for each container sandbox, there is one swap file for page fault based swap-in and the other REAP file for batch REAP swap-in. The swap files are dedicated for one sandbox and won’t be shared between sandboxes to mitigate potential secure vulnerability, these files are deleted when the sandbox terminates. 

Page fault based swap-in has different corresponding swap-out processes than REAP based swap-in, details are as below.

\subsubsection{Page fault based swap-in and swap-out}
This section describes the page fault based swap-in and corresponding swap-out.

The Serverless Platform may trigger an idle Warm Container’s hibernate process by sending the signal SIGSTOP to the Warm Container. After that, the Quark Container runtime’s Swapping Mgr in Figure 5 performs the following steps in order to swap-out memory.

\begin{enumerate}
\item \textbf{Pause Guest Applications}: The  Swapping Mgr pauses all the guest applications as a common SIGSTOP handler process. After that, the guest user application threads are blocked and won’t access any memory pages, so that the Swapping Mgr doesn’t need to handle any complex race condition and it simplifies the overall process;

\item \textbf{Walk through and modify guest applications page tables}: The Swapping Mgr walks through all the guest application page tables to identify the anonymous pages and do following work:
\begin{enumerate}
\item Mark each anonymous page's page table entry as Not-Present so that the later access to the page triggers a page fault;
\item Set the page table entry’s flags bit\#9, which is a customer bit, to indicate the page fault is due to page swap-out;
\item Put the page’s guest physical address in a hash table for page de-duplication in case there are multiple references for the same guest physical page address from multiple page tables.
\end{enumerate}
\item \textbf {Write the memory pages to the swap file}: The hibernate container has one swap file for each Quark Sandbox.  The Swapping Mgr enumerates the hash table and write the memory page to the swap file and then save the file offset of the memory page back to the hash table.
\item \textbf {Return the memory pages to host Linux OS}: The Swapping Mgr returns the swap-out pages to host Linux OS by calling syscall madvise().
\end{enumerate}

When a Hibernate Container is woken up, it resumes the execution of paused guest user applications. When the guest user application execution accesses a swap-out memory page, it triggers a page fault to swap-in the page. The page fault handling process is as below.

\begin{enumerate}
\item \textbf{Confirm whether the page fault is from swap-out page}: The Quark page handler checks the page table entry’s custom flag bit\#9. If the bit is set, it is a swap-out page and the swap-in process starts as step\#2.
\item \textbf{Load the memory from the swap file}: The Virtual CPU (vCPU)  of the page fault handler exits from guest mode to host mode and read the memory page from the swap file.
\item \textbf{Update page table entry}: The page table entry’s flag bit\#9 is cleared and the entry is also marked as Present so that no more page faults are triggered by the page.
\end{enumerate}

The page fault based swap-in cost is high. Its cost includes the following parts.
\begin{itemize}
\item \textbf{Page fault handling}: In a page fault handling process, the guest vCPU is transferred from guest user space to guest kernel space and all the general registers are stored in the main memory.
\item \textbf{Switch between guest mode and host mode}: The guest and host mode switch cost is high. It needs to store not only general registers but also float context in main memory. We observe about 15 microsecond latency for such a guest/host switch in our test environment
\item \textbf{Random page memory read from SSD}: Hibernate Container is tested with SSD disk. Although SSD’s random 4KB read throughput is much higher than HDD, it is still much lower than sequential batch read. From our test environment, the 4K page random read throughput is about 100MB/second while the sequential batch read throughput is more than 1GB/second.
\end{itemize}

From our observation, page fault based swap-in only loads 30\% to 90\% swap-out pages to process a user request. For example, in our experiment of Node.js Hello World hibernate test, a total  about 10 MB memory is swap-out and the user request processing only swaps-in about 4 MB memory. It is because the swap-out pages contain both memory pages for guest application initialization and request processing. And the application initialization has finished when inflating a Hibernate Container so that the related memory pages won’t be accessed and swap-in. Based on this observation and inspired by the REAP\cite{RCEP}, the Hibernate Container implemented a batch REAP pre-fetch swap-in mechanism.

\subsubsection{REAP swap-out and batch swap-in}

The main concept of Record-and-Prefetch (REAP) is to record the guest physical memory work set when processing an user request and batch prefetch them when next time wake up. It is an optimization of the page fault based swap-in mechanism. 

Compared with page fault based swap-out, REAP adds a Record process after the first time hibernation and wake-up. After the first time the container enter Hibernate state, the Record process does following steps:
\begin{enumerate}
\item \textbf{Sample user request}: The serverless platform sends a sample user request to trigger it to switch to the Hibernate Running state;
\item \textbf{Physical memory work set recording}: In the Hibernate Running state, the guest physical memory work set for the request process is loaded from the swap file through the page fault swap-in and the un-touch pages are still kept in the swap file;
\item \textbf{REAP hibernation}: After the sample user request process is done, the container returns to Woken-up state. The serverless platform sends the SIGSTOP signal to trigger the Woken-up Container hibernation. REAP swap-out process is triggered and its process is as below.
\begin{enumerate}
\item Pause all the guest user processes.
\item Walk through all the page tables to get all the active anonymous memory pages.
\item Records the guest physical memory page address in a scatter io vectors and save the memory pages in the REAP swap file in Figure 5 with batch write pwritev() syscall based on the io vectors.
\item Free the guest physical memory pages by calling syscall madvise().
\end{enumerate}
\end{enumerate}

We can see that REAP swap-out is different from the page fault based swap-out. 
\begin{itemize}
\item REAP swap-out won’t change the page table entries so that it won't trigger page fault;
\item REAP swap-out writes to a dedicated REAP swap file, which can accelerate the swap-in process through disk batch sequential read.
\end{itemize}

REAP swap-in process is simpler than page fault based swap-in. Its steps are as below.
\begin{enumerate}
\item Prefetch all memory pages from the REAP swap file with batch sequential read preadv() based on the scatter io vectors created in the REAP swap-out process.
\item Resume the guest application processes.
\end{enumerate}

REAP swapping process has higher performance than page fault based swap-in because of following optimization:
\begin{enumerate}
\item \textbf{No page fault overhead}: REAP swap-out doesn’t change the page table entry, so it won’t trigger the page fault when swap-in and the page fault handling and guest/host switch overhead is avoided.
\item \textbf{Batch file read}: REAP swap-in prefetches all memory pages in a batch read, which has much higher throughput than random reads.
\end{enumerate}

\subsection{File Backed Memory Sharing and Security Concerns}
Hibernate Container also cleans up file backed mmap memory with madvise() to return them to host Linux. Quark may share the file backed mmap memory between different containers with the Copy On Write. When the memory is shared, Hibernate Container deflate process doesn't need to clean up them as they may been using by other containers. Memory sharing may not only reduce container startup latency but also reduce the overall system memory footprint. This is quite an attractive technique for Serverless computing.

Unfortunately, in the multi-tenant Severless computing environment, when the file backed mmap memory is shared across different tenants, it results in secure risk such as cache side channel attack \cite{cacheattacks}. So the memory sharing is not recommended in the multi-tenant production environment \cite{FirecrackerRecommand}.

For applications running in a secure container, there are 2 major types of file backed memory might be shared cross containers:
\begin{itemize}
\item \textbf{User application program language runtime binary file}: User application is developed based on different program language runtime  binary file such as Node.js runtime and Python runtime. This type of files are memory mapped to user memory space and can be accessed directly by user application. It is risky to share them cross different tenants. 
\item \textbf{Secure container runtime binary file}: This is secure container run-time's executable binary and library files, such as the Linux guest kernel binary for Kata runtime. This type of files won't be mapped to user memory space and user applicatio can't access them directly, so the security risk is less than user application program language runtime binary. RunD \cite{RunD} is sharing Linux guest kernel in production environment to accelerate the cold startup and decrease memory footprint. 
\end{itemize}

Hibernate Container chooses to enable sharing of Quark runtime binary but disables sharing of program language runtime binary files. 

Runtime binary files sharing results in substantial request response latency reduction. For example, we evaluated Node.js runtime binary sharing in our test environment. When we enabled the Node.js binary memory sharing, the Hibernated Node.js hello-world container request response latency decreased from 25 ms to 11 ms. 

There are mitigations such as \cite{mitigating} \cite{CouldFlare} address the program language runtime sharing secure risk. Some mitigations are adopted in production deployment. For example Cloudflare worker \cite{CouldFlare} compromises secure risk of multi-tenant isolation based on  the V8 engine isolate. When the issue is solved, Hibernate Container may achieve better performance with enabling more file based memory sharing by adopting the mitigations.

\subsection{Hibernate Container Implementation}
Hibernate Container code-base is written in Rust programming language and runs as part of Quark secure container runtime environment. Quark runtime is a virtual userspace OS with 200K+ lines of Rust code. Hibernate Container functionality changes the critical code-path of Quark runtime such as virtual memory management, IO management and Virtual Machine Monitor (VMM). We implemented the Swapping Manager and Bitmap Allocator from scratch. The Swapping Manager has 780 lines of code and the Bitmap Allocator has 484 lines of code. We developed the Memory Reclaim Manager by modifying the Quark's file backed memory management and page tables management module and there are about 500 lines of code. There are also another about 300 lines of code change in Quark signal handling module and IO management for the hibernating process trigger.

\section{Evaluation}
In this section, we show the performance of a Hibernate Container using experiments. We conduct our experiments on a 1×12-core Intel(R) Core(TM) i7-8700K CPU @ 3.70GHz, 64GB RAM, PM981 NVMe Samsung 512GB SSD, running Ubuntu 20.04.4 Linux with 5.15.0-46-generic kernel. 

We use 2 set of micro-benchmarks to evaluate container hibernation's impact for request response latency and memory footprint. 
\begin{itemize}
\item Python benchmarks: We select a set of micro-benchmarks from Function Bench\cite{functionbench} to cover different process type, memory usage, process latency.  
\begin{itemize}
\item float-operation: This floating point arithmetic operations workload has small memory usage and process latency; 
\item video-processing: The Video Processing workload applies grayscale effect from the OpenCV library to a video input. It has more than 200 MB memory footprint and more than 1000 ms process latency.
\item image-processing: This program performs image transformation tasks using Python Pillow library. We choose 2 different size image files to evaluate data size impact for memory usage and process latency of same program. 
\end{itemize}
\item Program language runtime hello-world: We test Python, Node.js, Golang, Java hello-world programs to evaluate the behavior of different program language runtimes.
\end{itemize}

\subsection{User Request Response Latency}
This experiment demonstrates that the Hibernate Container user request response latency is lower than cold startup and the Woken-up Container has similar response latency to Warm Container. We also compare latency for the Page fault and REAP based swap-in.

As Hibernate Container has already started and in a keepalive state, instead of measuring application startup latency, we measure user application request/response end to end latency. Except the cold startup test, we run the micro-benchmarks in HTTP services and trigger the application processing with HTTP requests and then measure the HTTP response latency.

\includegraphics[scale=0.36]{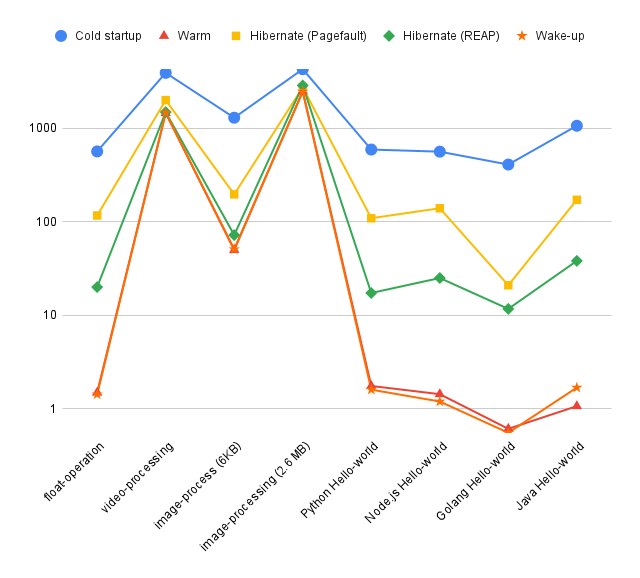}
\begin{center}
 Figure 6: Request Response Latency of Different Container State
\end{center}

In the experiment, we collect the latency from cold startup, Warm Container, Hibernate Container with page fault/REAP swap-in and Woken-up Container as below:
1. Cold startup: The process latency of a container startup and request handling without using HTTP request trigger;

2. Warm: The request response latency after the container is fully initialized; 

3. Hibernate: The request response latency of the first request after the container is Hibernated. We also collect latency for both Page fault and REAP swap-in mechanisms.

4. Woken-up: The request response latency of a Woken-up container.

Figure 6 shows the test results. From that, we can draw the following conclusions:
\begin{enumerate}
\item The Hibernate Container's request process latency is less than cold startup: For example, the Hibernate REAP request response latency takes 3\% (Python/Golang Hello-world) to 67\% (image-processing with 2.6 MB file) cold startup process latency. And the Hibernate REAP can save from 296 ms (Golang Hello-world) to 2407 ms (video-processing) over cold startup process latency.
\item The Woken-up Container requests process latency is almost similar to Warm Container. 
\item The request latency of Hibernate Container with page fault swap-in is more than REAP: In majority benchmarks, the REAP does better job than Page fault based swap-in. The only exception is from image-processing with 2.6 MB image file. But the difference is negligible. 
\end{enumerate}

The test results prove that Hibernate Container response latency is lower than cold startup and Woken-up Container process latency is similar as Warm Container. Additionally, Hibernate Container can benefit program language runtimes such as Python, Node.js, Golang and Java.

\subsection{Container Memory Consumption}

This experiment demonstrates that a Hibernate Container and its derive Woken-up Container consumes less memory than a Warm Container. We collect the benchmarks memory consumption Proportional Set Size (PSS) through Linux utility "pmap" for following state:
\begin{itemize}
\item Warm: The container processes a few user requests.
\item Hibernate: The container is transitioned to Hibernate state from Warm state.
\item Woken-up: The Hibernate Container is woken up by a user request.
\end{itemize}

As mentioned in Section 3.4 that Hibernate Container shares the Quark runtime binaries. So there is less memory PSS when there are more instances running. In our environment, we collect the PSS data with 10 running benchmark application instances. 

\includegraphics[scale=0.4]{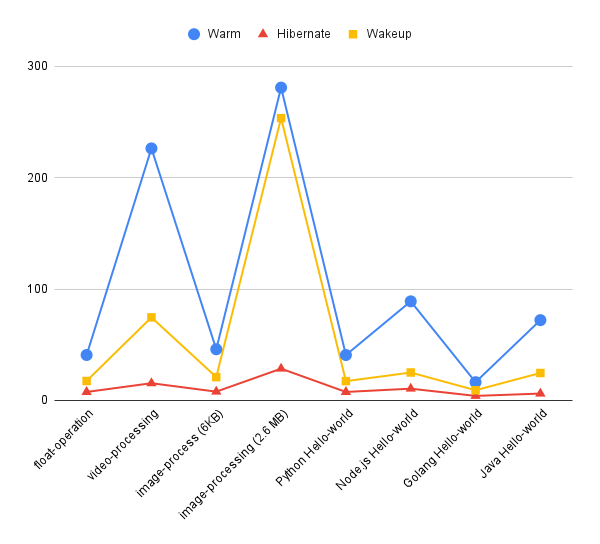}
\begin{center}
 Figure 7: Memory Consumption of Different Container State
\end{center}

We can draw the following conclusions from Figure 7:
\begin{enumerate}
\item The "Hibernate" state consumes much less memory than "Warm" state: For example, the "Hibernate" state consumes about 7\% (video-processing) to 25\% (Golang Hello-world) of "Warm" state memory. And the total memory saving ranges from 12 MB (of total 16MB Golang Hello-world) to 252 MB (of total 281MB image-processing with 2.6 MB file). 
\item The "Woken-up" state consumes less memory than "Warm" state: For example, the "Woken-up" state takes 28\% (Node.js Hello-world) to 90\% (image-processing with 2.6MB file) of "Warm" state memory. And the total memory saving ranges from 7MB  (of total 16 MB Golang Hello-world) to 151 MB  (of total 226 MB video-processing).
\end{enumerate}

The test result proves that a Hibernate Container and its derive Woken-up Container consume less memory than a Warm Container. And the memory saving is applicable for different program language runtime, includes Python, Node.js, Golang and Java.

And from both the response latency and memory test, we can conclude:
\begin{enumerate}
\item  Co-deployment of Hibernate Container and Woken-up Container can achieve higher deployment density than Warm Container.
\item Woken-up Container has less memory consumption but similar request latency to Warm container. It is good to covert a Warm Container to Woken-up Container through Hibernate Container as a keep-alive container when possible.
\end{enumerate}

\section{Related Work}
There has been several related work ongoing specifically work related to cold startup optimizations. Cold startup typically consists of 2 distinct activities: secure container runtime startup and guest user application startup. The following subsections describe various ongoing optimizations work for these two cold startup related activities.

\subsection{VM Based Secure Container Runtime Optimization}
A VM based secure container runtime startup includes Linux container environment (e.g. Cgroup, container network, container file system) setup and VM OS kernel boot up task. 

RunD \cite{RunD} strives to accelerate the container environment setup by pre-creating Cgroup, optimizing container rootfs mapping. RunD uses Kata template for reducing the per-microVM memory overhead as well as reduction in the startup latency.

Firecracker\cite{firecracker} introduced a light weight VMM to replace common VMM like QEMU or CloudHypervisor. Firecracker is optimized for serverless computing as it provides support for few devices only: support for a virtio network device and single block device type. This, in turn, results in optimization gains related to secure containers memory usage as well as sandbox startup latency.

Quark\cite{quark} and gVisor\cite{gVisor}, on the other hand, introduced both a new userspace OS Kernel and a new VMM dedicated for serverless workloads resulting in reduction in memory footprint and startup latency.

\subsection{Application Startup Optimization}
The key idea of above mentioned cold startup optimizations is to start the container from a state which is somewhat “closer” to a Warm Container. \cite{catalyzer}\cite{replayable} use checkpoint-restore technique provided by secure container runtime gVisor or JVM. \cite{Recycle} reuses a pending recycle Warm Container to host another container image. Based on checkpoint-restore (i.e.C/R), Catalyzer\cite{catalyzer} achieves init-less startup. There are more optimizations over C/R, e.g. REAP uses batch pre-fetch to accelerate the VMM image load. Sock\cite{sock} extends Zygote to start a new container by forking a helper container with pre-imported software packages to save the package import cost. Catalyzer introduces a Sandbox Fork (sfork) which fork from an existing Warm Container with a full application state and share the memory between the parent/child containers.

% It is important to highlight some of these cold startup optimization techniques still consume system memory while in an idle state. For example, the sandbox template of Catalyzer is kept in memory while in idle state and its memory consumption is similar to one Warm Container.

\section{Conclusion}
Low latency container startup time for a Serverless computing environment is key to improving the overall user experience. Besides Cold Startup and Warm Startup as part of the contemporary Serverless computing environment, this paper proposes the third startup mode: Hibernate Container. Hibernate Container is essentially a deflated Warm Container.

Our experiments demonstrate that Hibernate Container memory consumption is much lower than a Warm Container and its request response latency is less than a cold startup. At the time when Hibernate Container is woken-up, the Woken-up Container also consumes less memory than the Warm Container and has same request response latency at the same time. All of this results in much higher deployment density, lower request response latency, and appreciable improvements in the overall system performance.

\bibliographystyle{unsrt}
\bibliography{references}

\end{document}